\magnification \magstep1
\raggedbottom
\openup 2\jot
\voffset6truemm
\headline={\ifnum\pageno=1\hfill\else
\hfill {\it Quantized Maxwell Theory in a Conformally
Invariant Gauge} \hfill \fi}
\def\cstok#1{\leavevmode\thinspace\hbox{\vrule\vtop{\vbox{\hrule\kern1pt
\hbox{\vphantom{\tt/}\thinspace{\tt#1}\thinspace}}
\kern1pt\hrule}\vrule}\thinspace}
\rightline {DSF preprint 96/42}
\centerline {\bf QUANTIZED MAXWELL THEORY IN A}
\centerline {\bf CONFORMALLY INVARIANT GAUGE}
\vskip 1cm
\leftline {\it Giampiero Esposito}
\vskip 1cm
\leftline {Istituto Nazionale di Fisica Nucleare, Sezione di Napoli}
\leftline {Mostra d'Oltremare Padiglione 20, 80125 Napoli, Italy;}
\leftline {Dipartimento di Scienze Fisiche}
\leftline {Mostra d'Oltremare Padiglione 19, 80125 Napoli, Italy.}
\leftline {e-mail: esposito@napoli.infn.it}
\vskip 1cm
\noindent
{\bf Abstract.} Maxwell theory can be studied in a gauge which is
invariant under conformal rescalings of the metric, and first
proposed by Eastwood and Singer. This paper studies the corresponding
quantization in flat Euclidean 4-space. The resulting ghost operator
is a fourth-order elliptic operator, while the operator $\cal P$
on perturbations $A_{\mu}$ of the potential is a sixth-order elliptic
operator. The operator $\cal P$ may be reduced to a second-order 
non-minimal operator if a dimensionless gauge parameter tends to
infinity. Gauge-invariant boundary conditions are obtained 
by setting to zero at the boundary the whole set of $A_{\mu}$
perturbations, jointly with ghost perturbations and their normal 
derivative. This is made possible by the fourth-order nature of the
ghost operator. An analytic representation of the ghost basis
functions is also obtained.
\vskip 100cm
The recent attempts to quantize Euclidean Maxwell theory in quantum
cosmological backgrounds have led to a detailed investigation of the
quantized Maxwell field in covariant and non-covariant gauges on
manifolds with boundary [1--7]. The main emphasis
has been on the use of
analytic or geometric techniques to evaluate the one-loop semiclassical
approximation of the wave function of the universe, when magnetic
or electric boundary conditions are imposed. In the former case
one sets to zero at the boundary the tangential components $A_{k}$
of the potential (the background value of $A_{\mu}$ is taken to
vanish), the real-valued ghost fields $\omega$ and $\psi$
(or, equivalently, a complex-valued ghost 0-form $\varepsilon$),
and the gauge-averaging functional $\Phi(A)$:
$$
\Bigr[A_{k}\Bigr]_{\partial M}=0 \; ,
\eqno (1)
$$
$$
[\varepsilon]_{\partial M}=0 \; ,
\eqno (2)
$$
$$
\Bigr[\Phi(A)\Bigr]_{\partial M}=0 \; .
\eqno (3)
$$
In the electric scheme one sets instead to zero at the boundary
the normal component of $A_{\mu}$, jointly with the normal
derivative of the ghost and the normal derivative of $A_{k}$:
$$
\Bigr[A_{0}\Bigr]_{\partial M}=0 \; ,
\eqno (4)
$$
$$
\Bigr[{\partial \varepsilon}/ {\partial n}\Bigr]_{\partial M}=0 \; ,
\eqno (5)
$$
$$
\Bigr[{\partial A_{k}}/ {\partial n}\Bigr]_{\partial M}=0 \; .
\eqno (6)
$$
One may check that the boundary conditions (1)--(3) and
(4)--(6) are invariant under infinitesimal gauge transformations
on $A_{\mu}$, as well as under BRST transformations [5].

On the other hand, the gauge-averaging functionals studied in
Refs. [1--7] were not conformally invariant, although a conformally
invariant choice of gauge was already known, at the {\it classical}
level, from the work of Ref. [8]. It has been therefore our aim
to investigate the {\it quantum} counterpart of the conformally
invariant scheme proposed in Ref. [8], to complete the current
work on quantized gauge fields. For this purpose, we have
studied a portion of flat Euclidean 4-space bounded by 3-dimensional
surfaces. The vanishing curvature of the 4-dimensional background is
helpful to obtain a preliminary understanding of the quantum
operators, which will be shown to have highly non-trivial properties.
In our scheme, all curvature effects result from the boundary only.

In flat Euclidean 4-space, the conformally invariant gauge
proposed in Ref. [8] reads (hereafter $b,c=0,1,2,3$)
$$
\nabla_{b}\nabla^{b}\nabla^{c}A_{c}
=\cstok{\ }\nabla^{c}A_{c}=0 \; .
\eqno (7)
$$
If the classical potential is subject to an infinitesimal
gauge transformation
$$
{ }^{f}A_{b}=A_{b}+\nabla_{b}f \; ,
\eqno (8)
$$
the gauge condition (7) is satisfied by ${ }^{f}A_{b}$ if
and only if $f$ obeys the fourth-order equation
$$
\cstok{\ }^{2}f=0 \; ,
\eqno (9)
$$
where $\cstok{\ }^{2}$ is the $\cstok{\ }$ operator composed
with itself, i.e.
$\cstok{\ }^{2} \equiv g^{ab}g^{cd}\nabla_{a}\nabla_{b}
\nabla_{c}\nabla_{d}$. 

In the quantum theory via path integrals,
however, one performs Gaussian averages over gauge functionals
$\Phi(A)$ which ensure that well-defined Feynman Green's
functions for the $\cal P$ operator on $A_{b}$, and for the
ghost operator, actually exist [5,9]. This means that the
left-hand side of Eq. (7) is no longer set to zero. One defines
instead a gauge-averaging functional
$$
\Phi(A) \equiv \cstok{\ }\nabla^{b}A_{b} \; ,
\eqno (10)
$$
and the gauge-averaging term ${1\over 2\alpha}
[\Phi(A)]^{2}$, with $\alpha$ a dimensionless parameter,
is added to the Maxwell Lagrangian ${1\over 4}F_{ab}F^{ab}$.
A double integration by parts is then necessary to express 
the Euclidean Lagrangian in the form
${1\over 2}A_{b}{\cal P}^{bc}A_{c}$, where
$$
{\cal P}^{bc} \equiv -g^{bc}\cstok{\ }
+\biggr(1-{1\over \alpha}\cstok{\ }^{2}\biggr)
\nabla^{b}\nabla^{c} \; .
\eqno (11)
$$
The operator ${\cal P}^{bc}$ is a complicated sixth-order
elliptic operator, and it is unclear how to deal properly with
it for finite values of $\alpha$. However, in the limit
as $\alpha \rightarrow \infty$, it reduces to the following
second-order operator:
$$
P^{bc}=-g^{bc}\cstok{\ }+\nabla^{b}\nabla^{c} \; .
\eqno (12)
$$
This operator remains non-minimal, since the term 
$\nabla^{b}\nabla^{c}$ survives.
In this particular case, we still need to specify boundary
conditions on $A_{b}$ and ghost perturbations. For this 
purpose, we put to zero at the boundary the whole set of
$A_{b}$ perturbations:
$$
\Bigr[A_{b}\Bigr]_{\partial M}=0 \; \forall b=0,1,2,3 \; ,
\eqno (13)
$$
and we require invariance of (13) under infinitesimal gauge
transformations on $A_{b}$. This leads to (hereafter $\tau$
is a radial coordinate [1--4])
$$
[\varepsilon]_{\partial M}=0 \; ,
\eqno (14)
$$
$$
\Bigr[{\partial \varepsilon}/ \partial \tau 
\Bigr]_{\partial M}=0 \; .
\eqno (15)
$$
Condition (14) results from the gauge invariance of the 
Dirichlet condition on $A_{k}$, $\forall k=1,2,3$,
and condition (15) results from the gauge invariance of the
Dirichlet condition on $A_{0}$. Note that it would be
inconsistent to impose the boundary conditions (13)--(15)
when the Lorentz gauge-averaging functional is chosen, since
the corresponding ghost operator is second-order. 

When two boundary 3-surfaces occur, (14) and (15) lead to
$$
[\varepsilon]_{\Sigma_{1}}
=[\varepsilon]_{\Sigma_{2}}=0 \; ,
\eqno (16)
$$
$$
\Bigr[{\partial \varepsilon}/ \partial \tau 
\Bigr]_{\Sigma_{1}}
=\Bigr[{\partial \varepsilon}/ \partial \tau 
\Bigr]_{\Sigma_{2}}=0 \; .
\eqno (17)
$$
When Eq. (10) is used, and the ghost operator is hence
$\cstok{\ }^{2}$, the four boundary conditions (16) and
(17) provide enough conditions to determine completely
the coefficients $C_{1},...,C_{4}$ in the linear combination
$$
\varepsilon_{(\lambda)}=\sum_{i=1}^{4}C_{i}
\rho_{i_{(\lambda)}} \; ,
\eqno (18)
$$
where $\rho_{1},...,\rho_{4}$ are four linearly independent 
solutions of the fourth-order eigenvalue equation
$$
\cstok{\ }^{2}\varepsilon_{(\lambda)}=\lambda \;
\varepsilon_{(\lambda)} \; .
\eqno (19)
$$
We therefore find that, when the conformally invariant gauge
functionals (10) are used, the admissible boundary conditions
differ substantially from the magnetic and electric schemes 
outlined in Eqs. (1)--(3) and (4)--(6), and are conformally invariant
by construction (with the exception of Eq. (15)). 

Had we set to zero at the boundary $A_{k}$ ($k=1,2,3$) and the
functional (10), we would not have obtained enough boundary
conditions for ghost perturbations, since both choices
lead to Dirichlet conditions on the ghost. The boundary
conditions (13) are also very important since 
they ensure the vanishing of all boundary
terms resulting from integration by parts in the Faddeev-Popov
action. In the particular case when
the 3-surface $\Sigma_{1}$ shrinks to a point, which is relevant
for (one-loop) quantum cosmology [5], the boundary conditions
read (here $\Sigma$ is the bounding 3-surface)
$$
\Bigr[A_{b}\Bigr]_{\Sigma}=0 \; 
\forall b=0,1,2,3 \; ,
\eqno (20)
$$
$$
[\varepsilon]_{\Sigma}=0 \; ,
\eqno (21)
$$
$$
\Bigr[{\partial \varepsilon}/ \partial \tau \Bigr]_{\Sigma}
=0 \; ,
\eqno (22)
$$
jointly with regularity at $\tau=0$ of $A_{b}, \varepsilon$
and ${\partial \varepsilon \over \partial \tau}$. Many fascinating
problems are now in sight. They are as follows:
\vskip 0.3cm
\noindent
(i) To prove uniqueness of the solution of the classical 
boundary-value problem
$$
\cstok{\ }^{2}f=0 \; ,
\eqno (23)
$$
$$
[f]_{\Sigma_{1}}= [f]_{\Sigma_{2}}
=0 \; ,
\eqno (24)
$$
$$
\Bigr[{\partial f} / {\partial \tau} \Bigr]_{\Sigma_{1}}
=\Bigr[{\partial f} / {\partial \tau} \Bigr]_{\Sigma_{2}}
=0 \; .
\eqno (25)
$$
\vskip 0.3cm
\noindent
(ii) To study the quantum theory resulting from the operator
(11) for finite values of $\alpha$. Interestingly, the
Feynman choice $\alpha=1$ does not get rid of the sixth-order
nature of the operator ${\cal P}^{bc}$.
\vskip 0.3cm
\noindent
(iii) To evaluate the one-loop semiclassical approximation,
at least when ${\cal P}^{bc}$ reduces to the form (12) 
in the presence of 3-sphere boundaries. The ghost operator
is then found to take the form
$$
\eqalignno{
\cstok{\ }^{2}&={\partial^{4}\over \partial \tau^{4}}
+{6\over \tau}{\partial^{3}\over \partial \tau^{3}}
+{3\over \tau^{2}}{\partial^{2}\over \partial \tau^{2}}
-{3\over \tau^{3}}{\partial \over \partial \tau} \cr
&+{2\over \tau^{2}}\biggr({\partial^{2}\over \partial \tau^{2}}
+{1\over \tau}{\partial \over \partial \tau}\biggr)
{\;}_{\mid i}^{\; \; \; \mid i}
+{1\over \tau^{4}}\Bigr(
{\;}_{\mid i}^{\; \; \; \mid i}\Bigr)^{2}
\; . &(26)\cr}
$$
With a standard notation, we denote 
by $\mid$ the operation of covariant 
differentiation tangentially with respect to the 
3-dimensional Levi-Civita connection of the boundary.
If one expands the ghost perturbations on a family of
3-spheres centred on the origin as [1]
$$
\varepsilon(x,\tau)=\sum_{n=1}^{\infty}
\varepsilon_{n}(\tau)Q^{(n)}(x) \; ,
$$
the operator (26), jointly with the properties of scalar
harmonics, leads to the eigenvalue equation (cf. Eq. (19))
$$
\eqalignno{
\; & {d^{4}\varepsilon_{n}\over d\tau^{4}}
+{6\over \tau}{d^{3}\varepsilon_{n}\over d\tau^{3}}
-{(2n^{2}-5)\over \tau^{2}}{d^{2}\varepsilon_{n}
\over d\tau^{2}}
-{(2n^{2}+1)\over \tau^{3}}
{d\varepsilon_{n}\over d\tau} \cr
&+\left({(n^{2}-1)^{2}\over \tau^{4}}-\lambda_{n} \right)
\varepsilon_{n}=0 \; .
&(27)\cr}
$$
This equation admits a power series solution in the form
$$
\varepsilon_{n}(\tau)=\tau^{\rho}\sum_{k=0}^{\infty}
b_{n,k}(n,k,\lambda_{n})\tau^{k} \; .
\eqno (28)
$$
The values of $\rho$ are found by solving the fourth-order
algebraic equation
$$
\rho^{4}-2(n^{2}+1)\rho^{2}
+(n^{2}-1)^{2}=0 \; ,
\eqno (29)
$$
which has the four real roots $\pm (n \pm 1)$. 
Moreover, the only non-vanishing $b_{n,k}$ coefficients are 
of the form $b_{n,4k}$, $\forall k=0,1,2,...$, and are given
by (assuming that $b_{n,0}$ has been fixed)
$$
b_{n,l}={\lambda_{n} \; b_{n,l-4}\over
F(l,n,\rho)} \; , \; \forall l=4,8,12, ... \; ,
\eqno (30)
$$
where we have defined ($\forall k = 0,1,2,...$)
$$ 
\eqalignno{
F(k,n,\rho)& \equiv (\rho+k)(\rho+k-1)(\rho+k-2)(\rho+k-3)
+6(\rho+k)(\rho+k-1)(\rho+k-2) \cr
&-(2n^{2}-5)(\rho+k)(\rho+k-1)
-(2n^{2}+1)(\rho+k)
+(n^{2}-1)^{2} \; .
&(31)\cr}
$$
As far as we can see, the solution (28) can be expressed
in terms of Bessel functions and of a new set of special
functions (cf. Sec. 3.5 of Ref. [10]).
\vskip 0.3cm
\noindent
(iv) To include the effects of curvature. As shown in Ref. [8],
if the background 4-geometry is curved, with Riemann tensor
$R_{\; \; bcd}^{a}$, the conformally invariant 
gauge-averaging functional reads (cf. Eq. (10))
$$
\Phi(A) \equiv \cstok{\ }\nabla^{b}A_{b}
+\nabla_{c} \left[\Bigr(-2R^{bc}+{2\over 3}R g^{bc}\Bigr)
A_{b}\right] \; .
\eqno (32)
$$
It would be interesting to study the (one-loop) quantum theory,
at least when $\alpha \rightarrow \infty$, on curved backgrounds
like $S^{4}$, which is relevant for inflation [5], or
$S^{2} \times S^{2}$, which is relevant for the bubbles picture
in Euclidean quantum gravity, as proposed in Ref. [11].

To our knowledge, the form (11) of the differential operator 
on perturbations of the electromagnetic potential in the
quantum theory, the boundary conditions (13)--(15), and the
analytic solution (28)--(31) for ghost basis functions are
entirely new. Thus, quantization via path integrals in
conformally invariant gauges possesses some new peculiar 
properties, which are now under investigation for the first
time. This, in turn, seems to add evidence in favour of
Euclidean quantum gravity having a deep influence on current
developments in quantum field theory [5].

The author is indebted to A. Yu. Kamenshchik and G. Pollifrone
for scientific collaboration on Euclidean Maxwell theory and
Euclidean quantum gravity over many years.
\vskip 1cm
\leftline {\bf References}
\vskip 1cm
\item {[1]}
G. Esposito, Class. Quantum Grav. 11 (1994) 905.
\item {[2]}
G. Esposito and A.Yu. Kamenshchik, Phys. Lett. B 336 (1994) 324.
\item {[3]}
G. Esposito, A.Yu. Kamenshchik, I.V. Mishakov and G. Pollifrone,
Class. Quantum Grav. 11 (1994) 2939.
\item {[4]}
G. Esposito, A.Yu. Kamenshchik, I.V. Mishakov and G. Pollifrone,
Phys. Rev. D 52 (1995) 2183.
\item {[5]}
G. Esposito, A.Yu. Kamenshchik and G. Pollifrone, {\it Euclidean
Quantum Gravity on Manifolds with Boundary} 
(Kluwer, Dordrecht, 1997).
\item {[6]}
D.V. Vassilevich, J. Math. Phys. 36 (1995) 3174.
\item {[7]}
D.V. Vassilevich, Phys. Rev. D 52 (1995) 999.
\item {[8]}
M. Eastwood and M. Singer, Phys. Lett. 107 A (1985) 73.
\item {[9]}
B.S. DeWitt, in {\it Quantum Gravity 2, a Second Oxford Symposium},
eds. C.J. Isham, R. Penrose and D.W. Sciama
(Oxford, Clarendon Press, 1981).
\item {[10]}
W. Magnus, F. Oberhettinger, and R. P. Soni, {\it Formulas
and Theorems for the Special Functions of Mathematical
Physics} (Springer-Verlag, Berlin, 1966).
\item {[11]}
S.W. Hawking, Phys. Rev. D 53 (1996) 3099.

\bye